\makeatletter
\@ifundefined{@parse@version@dash}{%
\def\@parse@version#1{\@parse@version@0#1}
\def\@parse@version@#1/#2/#3#4#5\@nil{%
\@parse@version@dash#1-#2-#3#4\@nil}
\def\@parse@version@dash#1-#2-#3#4#5\@nil{%
  \if\relax#2\relax\else#1\fi#2#3#4 }
}{}
\makeatother

\documentclass[twocolumn,secnumarabic,amsmath,amssymb,aps,pra,superscriptaddress,eqsecnum,longbibliography]{revtex4-2}
\usepackage{graphicx}
\usepackage{bm}
\usepackage{xcolor}
\usepackage[normalem]{ulem}
\usepackage{braket}
\usepackage{algpseudocode}
\usepackage{algorithm}
\usepackage{booktabs}
\usepackage{physics}
\usepackage{soul}
\usepackage{tikz-imagelabels}
\usepackage{float}

\usepackage{here}

\usepackage{amsthm}
\theoremstyle{definition}

\newtheoremstyle{mystyle}
    {}
    {}   
    {\itshape}
    {}            
    {\bfseries}
    {.}                  
    { }
    {\thmname{#1}\thmnumber{ #2}\thmnote{ (#3)}}
\theoremstyle{mystyle}

\usepackage[,
colorlinks=true,
urlcolor=blue,
citecolor=blue,
linkcolor=blue,
hyperfootnotes=false]{hyperref}

\begin{document}
\title{Automatic Structural Search of Tensor Network States \\ including Entanglement Renormalization}

\author{Ryo Watanabe}
\email{u293494e@ecs.osaka-u.ac.jp}
\affiliation{Graduate School of Engineering Science, Osaka University, 1-3 Machikaneyama, Toyonaka, Osaka 560-8531,
Japan}

\author{Hiroshi Ueda}
\email{ueda.hiroshi.qiqb@osaka-u.ac.jp}
\affiliation{Center for Quantum Information and Quantum Biology, Osaka University, Toyonaka, Osaka 560-0043, Japan}
\affiliation{Computational Materials Science Research Team, RIKEN Center for Computational Science (R-CCS), Kobe, Hyogo 650-0047, Japan}

\begin{abstract}
Tensor network (TN) states, including entanglement renormalization (ER), can encompass a wider variety of entangled states.
When the entanglement structure of the quantum state of interest is non-uniform in real space, accurately representing the state with a limited number of degrees of freedom hinges on appropriately configuring the TN to align with the entanglement pattern.
However, a proposal has yet to show a structural search of ER due to its high computational cost and the lack of flexibility in its algorithm.
In this study, we conducted an optimal structural search of TN, including ER, based on the reconstruction of their local structures with respect to variational energy.
Firstly, we demonstrated that our algorithm for the spin-$1/2$ tetramer singlets model could calculate exact ground energy using the multi-scale entanglement renormalization ansatz (MERA) structure as an initial TN structure.
Subsequently, we applied our algorithm to the random XY models with the two initial structures: MERA and the suitable structure underlying the strong disordered renormalization group. 
We found that, in both cases, our algorithm achieves improvements in variational energy, fidelity, and entanglement entropy. 
The degree of improvement in these quantities is superior in the latter case compared to the former, suggesting that utilizing an existing TN design method as a preprocessing step is important for maximizing our algorithm's performance.

\end{abstract}

\date{\today}

\maketitle

\section{Introduction}\label{sec:Intro}
Developing numerical methods to analyze non-uniformly or intricately entangled quantum states is crucial for comprehending the physical properties of realistic quantum systems, including chemical and disordered systems.
For instance, Anderson explained phenomena such as spin diffusion and conduction in lattices with impurities or defects using a simple model incorporating disorder~\cite{PhysRev.109.1492}.
Furthermore, many-body localization has attracted significant interest within the realm of strongly disordered quantum many-body systems~\cite{basko_metalinsulator_2006,PhysRevB.105.224203}.

To address these problems, the strong disorder renormalization group (SDRG)~\cite{PhysRevLett.43.1434, PhysRevB.22.1305} has been developed, capable of approximating the qualitative properties of the ground state entanglement.
Subsequently, the density matrix renormalization group (DMRG) for random systems~\cite{doi:10.1143/JPSJ.65.895} was introduced as a variational method that can be systematically improved and extends beyond the real-space RG based on perturbation theory.

Concurrently, SDRG was combined with TN methods for extensive numerical calculations, such as in the case of higher dimensional systems.
The combined methods were specifically designed to be applicable in the context of tree tensor network (TTN)~\cite{PhysRevB.60.12116,PhysRevB.104.134405,PhysRevB.89.214203} and entanglement renormalization (ER)~\cite{PhysRevB.96.155136}.
The TTN state~\cite{PhysRevA.74.022320,PhysRevB.80.235127,PhysRevB.82.205105}, a generalization of matrix product states (MPS)~\cite{10.5555/2011832.2011833,PhysRevLett.123.170504} known as the variational state of DMRG~\cite{PhysRevLett.69.2863,schollwock_density-matrix_2011}.
ER, was initially introduced in multi-scale entanglement renormalization ansatz (MERA)~\cite{PhysRevLett.101.110501,PhysRevB.79.144108}, a generalization of TTN, to describe the logarithmic divergence of the entanglement entropy in one-dimensional (1D) quantum systems, were the theoretical underpinnings for this approach.

In the variational approach to non-uniform systems using tensor networks (TNs)~\cite{ORUS2014_ann.phys,Orus2014_Eur.Phys.J.B,ONU2022TN}, as described above, tensor elements are optimized with the whole TN structure fixing.
However, optimizing the structure to align with the entanglement layout within the target quantum state is essential to obtain more accurate solutions with a fixed degrees of freedom ~\cite{PhysRevB.101.195134}.
This optimization involves combinatorial optimization and poses a significant challenge, given that the number of network patterns in the TN increases exponentially with the number of tensors.
Initially pioneered in TN calculations for molecular systems in quantum chemistry~\cite{PhysRevB.68.195116,rissler_measuring_2006,10.1063/1.4798639}, the structural optimization with TTNs has recently exhibited attention based on the same principles in the context of both regular and random quantum spin system~\cite{okunishi2023ptep,PhysRevResearch.5.013031,hikihara2024visualization}.

Extending these ideas to incorporate ER into TN (ER-TN) represents a sure foothold in advancing TN methods for non-uniform systems, although it is acknowledged that this process is challenging.
For instance, we examine an approach discussed in Ref.~\cite{PhysRevResearch.5.013031}, which explores the reconstruction of local structures, including the bipartition regions of given TTNs.
It aims to minimize the entanglement entropy for the bipartition while simultaneously performing a two-tensor update with the DMRG.
Unfortunately, this approach is difficult to adapt to ER-TN states due to the inner loops existing in the networks' layout.
The underlying challenge arises from the complexity involved in evaluating entanglement entropy within these TN states, attributable to the involvement of a multitude of tensors.

In this study, we introduce a scheme for an automated optimal structural search focusing on the reconstruction of their local structures of ER-TN and benchmark it in the spin $S=1/2$ quantum spin systems.
By imposing a fixed rank on each tensor, we restrict the number of possible structures with of two adjacent tensors
Although our method directly references a cost function relevant to the entire TN structure, updating and evaluating local tensors can lead to being trapped in local minima.
To address this issue, we incorporate ideas of sampling from the Gibbs distribution and replica exchange~\cite{doi:10.1143/JPSJ.65.1604,PhysRevE.100.043311}, well-known heuristic methods commonly used to solve classical combinatorial optimization problems.

Our proposed scheme extends to general isometric TN states, potentially benefiting the advancement of quantum computing and quantum information processing~\cite{PhysRevX.10.041038,PRXQuantum.5.010308,PRXQuantum.5.010308,doi:10.1126/sciadv.adk4321}.
It becomes particularly vital in the era of noisy intermediate-scale quantum (NISQ) computing devices, where even small-scale investigations can yield significant insights~\cite{Cerezo2021,RevModPhys.94.015004,PhysRevResearch.6.023009}.

The organization of this paper is structured as follows:
Section \ref{sec:Introduction of RG} provides an overview of ER-TN and its variational algorithms. 
Section \ref{sec:Method}, elaborates on our proposed scheme for optimizing structures of the ER-TN.
Section \ref{sec:Numerical Simulation} presents benchmark results for $S = 1/2$ Heisenberg tetramer chains~\cite{doi:10.7566/JPSJ.87.123703} and disordered XY chains.
For the latter models, we applied a previously established method~\cite{PhysRevB.96.155136} to determine the sub-optimal structure and compared the results.
Finally, Section~\ref{sec:Conclusion} summarizes our findings and explores prospective avenues for future research, emphasizing practical strategies for applying our method to large-scale systems.

\section{Tensor Network States with Entanglement Renormalization}\label{sec:Introduction of RG}
Here, we briefly review of the ER-TN, a type of isometric TN, and its optimization algorithms.
Fig.~\ref{fig:iso-TN, isometric_condition} (a) presents the 1D binary MERA network as the simplest example of such states.
The network of Fig.~\ref{fig:iso-TN, isometric_condition} (a) is constructed using three types of tensors: a four-leg disentangler $u$, a three-leg isometry $v$, and a two-leg top tensor $t$.
In this research, we focus on ER-TNs shown in Fig.~\ref{fig:iso-TN, isometric_condition} (a) and its variants; however, any general TN composed of isometric tensors, regardless of the number of legs on each tensor, is classified as an ER-TN.

These tensors satisfy the following properties called isometric conditions
\begin{eqnarray}
  \sum_{cd} \left(u^{ab}_{cd}\right)^* u^{a'b'}_{cd} & = & \delta_{aa'} \delta_{bb'}, \label{eq:isometric_u_1}\\
  \sum_{ab} \left(u^{ab}_{cd}\right)^* u^{ab}_{c'd'} & = & \delta_{cc'} \delta_{dd'}, \label{eq:isometric_u_2}\\
  \sum_{ab} \left(v^{c}_{ab}\right)^* v^{c'}_{ab} & = & \delta_{cc'}, \label{eq:isometric_w}\\
  \sum_{ab} \left(t_{ab}\right)^* t_{ab} & = & 1~,
\end{eqnarray}
where $\delta_{\alpha\alpha'}~(\alpha \in \{a,b,c,d \})$ is the Kronecker's delta.
These conditions are schematically represented in Fig.~\ref{fig:iso-TN, isometric_condition} (b).
The TN state $\ket{\Psi}$ is always normalized, as the isometric conditions indicate, ensuring $\braket{\Psi}=1$.

A central concept of ER is to improve the accuracy of coarse-graining of $v$ by employing $u$ to disentangle local entanglement.
Owing to these properties, ER can effectively and efficiently capture long-range entanglement~\cite{Kuwahara2020,McMahon2020}.
However, introducing disentanglers in networks leads to the formation of loop structures, making it difficult to bipartite and analyze tensor networks.

\begin{figure}[H]
  \centering
  \includegraphics[clip, width=3.3in]{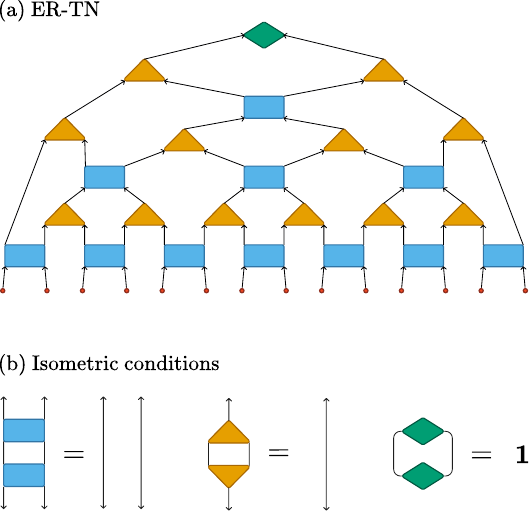}
  \caption{Schematic diagrams of (a) an example of isometric TN states with ER for the 14-site system, where blue, orange, and green tensors indicate disentanglers, isometries, and top tensors and red circles represent bare spins. Arrows indicate the direction of renormalization. (b) Isometric conditions of disentangler, isometry, and top tensors.}
  \label{fig:iso-TN, isometric_condition}
\end{figure}

Let us explore the optimization problem of a tensor that constitutes the TN state to search the ground state of the Hamiltonian $\mathcal{H}$.
In this paper, the Hamiltonian consists of the sum of the two-body interactions $h_{ij}$ between $i$-th and $j$-th sites, namely $\mathcal{H} = \sum_{i < j} h_{ij}$.
Therefore, to minimize the variational energy $E = \expval{\mathcal{H}}{\Psi}$, the goal is to optimize a tensor $v$ while keeping the remaining tensors fixed.
The energy $E$ depends on $v$ and its conjugate tensor $v^{\dagger}$ quadratically as
\begin{eqnarray}
  E(v) = \sum_{abca'b'c'} \sum_{\langle i<j \rangle} \left(v^{c}_{ab}\right)^* \left[\mathcal{E}_{ij}\right]^{abc}_{a'b'c'} v^{c'}_{a'b'} + C~,
\end{eqnarray}
where $\langle i<j \rangle$ refers to the set of site pairs specifying the two-body interactions $h_{ij}$ contributing to the optimization of $v$.
$\mathcal{E}_{ij}$ is the tensor obtained by cutting out $v$ and $v^{\dagger}$ from the tensor network to evaluate $e_{ij} = \expval{h_{ij}}{\Psi}$,
and $C=\sum_{\overline{\langle i<j \rangle}} e_{ij}$ with $\sum_{i<j}=\sum_{\langle i<j\rangle} + \sum_{\overline{\langle i<j\rangle}}$ being a constant term with respect to the update of $v$.
Regrettably, there is currently no known algorithm for optimizing $v$ in quadratic form while maintaining isometric constraints.

Therefore, as an alternative, we linearize the cost function and solve a linear problem.
In this procedure, we introduce
\begin{eqnarray}
  \tilde{E}(v) & \equiv & \sum_{abc} \left[ \Upsilon_v \right]^c_{ab} v^c_{ab} ~, \\
  \left[\Upsilon_v\right]^{c'}_{a'b'} & \equiv & \sum_{abc} \sum_{\langle i<j \rangle}  \left(v^{c}_{ab}\right)^* \left[\mathcal{E}_{ij}\right]^{abc}_{a'b'c'} ~,
\end{eqnarray}
where $\Upsilon_v$ is called the environment tensor of $v$.
Thus, the partial derivative of $E(v)$ concerning $v$ is denoted by
\begin{eqnarray}
  D_v = 2 \partial_v \tilde{E} = 2 \Upsilon_v~.
\end{eqnarray}
It's worth noting that when calculating $D_v$, one can effectively utilize automatic differentiation with the latest advanced tensor operation libraries~\cite{PhysRevX.9.031041,Geng_2022,Luchnikov_2021} instead of performing the contraction of the environment tensor of $v$.

Two main algorithms are commonly employed to achieve a unique global minimization of $\tilde{E}$: the algebraic-based optimization proposed by Evenbly and Vidal~\cite{PhysRevB.79.144108}, and the Riemannian optimization on the Stiefel manifold~\cite{10.21468/SciPostPhys.10.2.040,Luchnikov_2021}.

In the former algorithm, a singular value decomposition (SVD)
\begin{equation}
  \left[ \Upsilon_v \right]^{c}_{ab} \stackrel{\rm{SVD}}{=} \sum_{c'} V_{cc'}S_{c'}W^*_{c'ab}~,
\end{equation}
where $V, W$ are isometries, and $S$ is a positive real vector,
is applied to the environment tensor $\Upsilon_v$.
The tensor $v$ is always updated as follows
\begin{equation}
  [v]^c_{ab} \leftarrow -\sum_{c'} V_{cc'}W^*_{c'ab} ~.
\end{equation}
to maximize the absolute value of the energy $\tilde{E}(v)$.
To ensure that this update consistently converges to global energies, namely to minimize $\tilde{E}(v)$, the Hamiltonian should be redefined as $\mathcal{H}_{\gamma} = \mathcal{H} - \gamma I$, where $\gamma$ must be sufficiently large so that $\mathcal{H}_{\gamma}$ is negative-definite.
Nevertheless, since the optimization step size is scaled by $\gamma^{-1}$~\cite{10.21468/SciPostPhys.10.2.040}, so $\gamma$ should be chosen to be as small as practically possible.

Regarding the latter algorithm, the definite difference point with the algebraic-based one is that gradient-based methods can update each tensor slightly with the learning rate:
\begin{equation}
  v \leftarrow v - \eta g(v) ~,
  \label{eq:RiemannianOptimization}
\end{equation}
where $\eta$ is the learning rate and $g(v)$ is the Riemannian gradient of derivative $D_v$.
After the update described in Eq.~\eqref{eq:RiemannianOptimization}, retracting the updated tensor $v$ is crucial to ensure its alignment with the Stiefel manifold.

Hence, upon obtaining the partial derivatives $\{ D_v \}$ of all tensors, which can be computed simultaneously using automatic differentiation for the variational energy $E$, we can update all tensors concurrently.
Riemannian optimization methods primarily converge faster than previous methods when equipped with properly chosen hyperparameters because they eliminate the need to redefine the Hamiltonian with $\gamma$.
Various gradient-based optimization techniques, such as nonlinear conjugate gradient and quasi-Newton algorithms, are available, providing additional optimization options.

\section{Method}\label{sec:Method}
\begin{figure*}[!t]
  \begin{center}
    \includegraphics[width=0.9\textwidth]{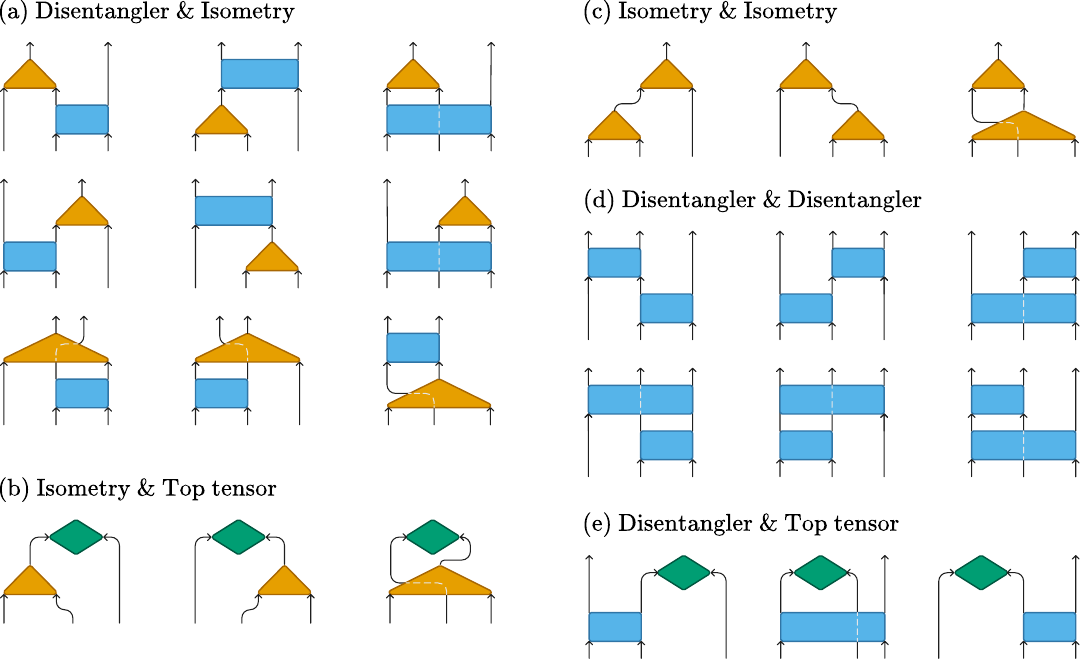}
    \caption{Schematic diagrams of defining local structures in the TN, including $u$, $v$, and $t$, representing (a) $\{u, v\}$, (b) $\{v, t\}$, (c) $\{v, v\}$, (d) $\{u, u\}$, and (e) $\{u, t\}$ type respectively.}
    \label{fig:local_structures}
    \end{center}
\end{figure*}
We will now describe our algorithm for the structural search of ER-TN.
In order to focus on the effects of this algorithm, we set bond dimensions at each leg of tensors being $\chi = 2$.
The fundamental concept of our algorithm lies in the reconstruction of pairs of tensors~\cite{PhysRevResearch.5.013031} in a given ER-TN.

An abstract of procedures of our algorithm is outlined in Table.~\ref{Table_I}.
\begin{table}[H]
\caption{Automatic structural search for the ER-TN.}
\label{Table_I}
\begin{center}
\begin{tabular}{ll}
\hline
\hline
1. & Select a pair of adjacent tensors within a given TN\\
   & (see Appendix.~\ref{appendix: our_algorithm}).\\
2. & Update two tensors for each possible TN structure under \\ 
   & fixed bond dimensions to minimize the variational energy. \\
3. & Adopt a local structure stochastically using Eq.~\eqref{eq:gibbs distribution}. \\
4. & Iterate the steps 1-3 until the entire network is updated. \\
5. & Update all tensors with respect to the energy while keeping \\
   & the whole TN structure fixed. \\
\hline
\hline
\end{tabular}
\end{center}
\end{table}
Before implementing our algorithm, we must define the local structures according to the given TN.
For example, in the ER-TN composed of only $u$, $v$, and $t$, all local structures depicted in Fig.~\ref{fig:local_structures} meet the isometric conditions for each of the five types: $\{u, v\}$, $\{v, t\}$, $\{v, v\}$, $\{u, u\}$ and $\{u, t\}$.
Note that if we use other isometric tensors, such as a three-site isometry, we have to consider new local structures with them.

\begin{figure}[H]
  \centering
  \includegraphics[clip, width=2.8in]{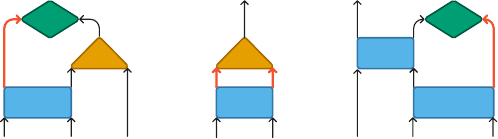}
  \caption{The examples of the pair of tensors, indicated by bold red edge $p$, to which our method cannot be applied. We have to update $f_p$ to $1$ before the step 1.}
  \label{fig:local loop strucure}
\end{figure}

After that, in the first step of Table.~\ref{Table_I}, our algorithm determines a pair of tensors.
Here, we assume the given TN has $N_{\rm T}$ tensors and $N_{\rm E}$ edges connecting for two tensors.
In our algorithm, one pair of tensors is specified by the two-dimensional vector $e_p$ that stores two integers $(n,n'\neq n) \in [1,N_{\rm T}]^{\otimes 2}$ identifying the two tensors connected by the $p$-th edge with $p\in[1,N_{\rm E}]$.
We define the integer $f_p \in \{ 0, 1 \}$ to indicate the status (flag) of updates to the local structure associated with its corresponding $e_p$ during the step 4.
If $f_p = 0$, the update of adjacent two tensors specified by $e_p$ is still pending, and if $f_p = 1$, it indicates that the update has been completed.
Initially, we set the $f_p$ to 1 for $p$ corresponding to bare spins and $0$ for the others.
Moreover, before each choice of tensor's pair, we ensure to also set the $f_p$ to 1 in case the two tensors mentioned by $e_p$ form loops, as depicted in Fig.~\ref{fig:local loop strucure}, because these tensors cannot be reconnected within prepared manners in Fig.~\ref{fig:local_structures}.

Additionally, we introduce the integer $d_p \in [1,N_{\rm T}]$ representing the distance, that is, the minimum number of tensors required to pass from the top tensor $t$ to edge $p$ along the structure of TN.
Given these values, our algorithm selects the $p$-th edge with largest $d_p$ among the set of $e_p$ where $f_p=0$.
This strategy, which refers to $d_p$, is pivotal in ER as it efficiently disentangles a region near the physical spaces.
Furthermore, our algorithm prioritizes the adjacent local structures specified by $e_p$ to the one selected in previous step 1 as explained in Appendix.~\ref{appendix: our_algorithm}.
If more than one candidate passes the above selections, our scheme selects one randomly from among them.

In the second step of Table.~\ref{Table_I}, our algorithm performs the variational optimization concerning energy for local tensors across all structural configurations to assess a better structure for the target state.
Specifically, we firstly initialize two tensors, which are specified in the first step, with $\chi=2$ to satisfy the below conditions
\begin{equation}
    \{u^{ab}_{cd}\} = \begin{pmatrix}
    1 & 0 & 0 & 0 \\
    0 & 1 & 0 & 0 \\
    0 & 0 & 1 & 0 \\
    0 & 0 & 0 & 1
    \end{pmatrix},~
    \{v^{a}_{bc}\} = \begin{pmatrix}
    1 & 0 \\
    0 & 1 \\
    0 & 0 \\
    0 & 0
    \end{pmatrix},~
    \{t_{ab}\} = \begin{pmatrix}
    1 \\
    0 \\
    0 \\
    0 
    \end{pmatrix},
\end{equation}
respectably.
In these equations, $\{ u^{ab}_{cd }\}$ and $\{ v^{a}_{bc}\}$ represent tensors $ u^{ab}_{cd}$ and $ v^{a}_{bc}$ in matrix form, where the upper indices correspond to columns, and the lower indices correspond to rows, and $\{ t_{ab} \}$ is vector form of $t_{ab}$.
Then, we update them through the variational calculation.
For the update, we used the Adam algorithm with the learning rate $\eta$ using the protocol of the Riemannian optimization~\cite{kingma2017adam}, as presented in Sec.~\ref{sec:Introduction of RG}.
In our algorithm, we adopt a strategy where $\eta$ is progressively reduced from its initial value, $\eta_{\text{init}}$, to its final value, $\eta_{\text{end}}$.
This reduction is defined as
\begin{equation}\label{eq:learning rate}
    \eta \leftarrow \eta_{\text{init}} - (l-1)\frac{\eta_{\text{init}} - \eta_{\text{end}}}{N_L-1}~,
\end{equation}
where $N_L$ represents the total number of iterations, and $ l \in [1, N_L] $ is the current iteration number.
Furthermore, to ensure efficiency in our algorithm, our algorithm terminates the updating routine at iteration $l$, if the difference $|E_{l} - E_{l-1}|$ is less than $\delta$, a threshold indicating that the variational energy has converged to the desired accuracy.
We will discuss specific values, $N_L$, $\{ \eta_{\text{init}}, \eta_{\text{end}} \}$, and $\delta$ in Sec.~\ref{sec:Numerical Simulation}.

When selecting a new configuration during the third step in Table.~\ref{Table_I}, we incorporated the heat bath method, which operates within the space of variational energy solutions for each structural candidate, together with the Boltzmann distribution to avoid the local minima with respect to the TN structures:
\begin{equation}\label{eq:gibbs distribution}
  P_i = \frac{\exp(-\beta E_i)}{\sum_j \exp(-\beta E_j)} ~,
\end{equation}
where $\beta$ represents the inverse temperature and $E_i$ denotes the variational energy of the structural arrangement $i$.
In Eq.~\eqref{eq:gibbs distribution}, an infinite $\beta \rightarrow \infty$ implies that we always select optimal structure.

After completing the step 4, at the fifth step of Table.~\ref{Table_I}, we update tensors globally under fixing the obtained TN structures.
We will further explain its specific hyperparameters in Sec.~\ref{sec:Numerical Simulation}.

If we repeat the entire process outlined in Table.~\ref{Table_I}, we could also integrate the idea of replica exchange methods~\cite{PhysRevLett.57.2607} at the end of fifth step in Table.~\ref{Table_I} to address the issue of local optimal solution.
It is necessary to first create $N_R$ replicas with different finite values of $\beta$, evenly distributed within a specific range $[\beta_{\min}, \beta_{\max}]$, and then apply our algorithm to all replicas.
Following this, using the Metropolis exchange method, we exchange adjacent replicas with acceptance probabilities
\begin{eqnarray}\label{eq:replica exchange}
  P = \min \left(1, \exp(- \Delta S) \right)~,
\end{eqnarray}
determined by $\Delta S = (\beta_{r+1} - \beta_{r}) ( E_{r+1} - E_{r} )$, where $r \in [1, N_R-1]$ indicate the index of replica.

Furthermore, strategies for pruning unnecessary structures can be incorporated to enhance the efficiency of the search algorithm.
At each step 3, if the energy difference between the structure with the lowest energy and one of the second lowest energy exceeded $\delta_{\textrm{structure}}$, we deterministically selected the structure associated with the lowest energy as the optimal one in each replica, even in the presence of a heat bath with a finite $\beta$.
At each replica exchange, if the energy difference between replicas with the lowest energy and those with the second lowest energy exceeded $\delta_{\textrm{replica}}$, we assigned the best structure across all replicas.

\section{Numerical Simulation}\label{sec:Numerical Simulation}
\begin{figure*}
  \begin{center}
    \includegraphics[width=1.0\textwidth]{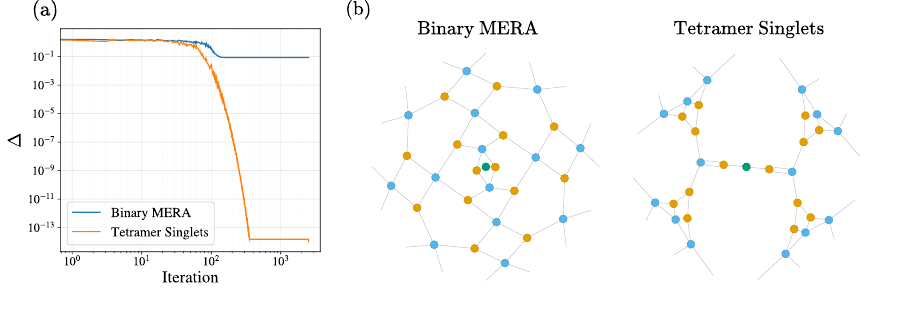}
    \caption{(a) The iteration dependence of relative errors $\Delta$, for spin $S=1/2$ tetramer model, with $L=4$ and $J'/J = 0.0$, across binary MERA and tetramer singlets structures in (b). In two network diagrams, blue, orange, and green circles represent disentangler, isometry, and top tensor, respectively. The tetramer singlets structure is determined manually by the definition of the Hamiltonian Eq.~\eqref{eq:tetramer_ham}. It is divided between bare spin groups consisting of a tetramer singlet.}
    \label{fig:test_tetramer}
    \end{center}
\end{figure*}
In this section, we will show the benchmark calculations for our algorithm, focusing on the ground state search of two quantum many-body systems.
The first system is the spin-$1/2$ tetramer model~\cite{doi:10.7566/JPSJ.87.123703}, and it has been established that the exact ground state in the tetramer-singlet phase can be obtained from rebuilding TN from 1D binary MERA with $\chi=2$, making it suitable for verifying the adequacy of our method.
The second is the 1D random XY model as a main target of this paper, for which examples of ER-TN calculations incorporating SDRG have already been reported~\cite{PhysRevB.96.155136}. 
We demonstrate that applying our structural optimization to the ER-TNs generated by SDRG enables us to approach more accurate solutions with respect to energy, fidelity, and entanglement entropy without changing the bond dimension of tensors.

In this research, we used the TensorNetwork library~\cite{roberts2019tensornetwork} for TN calculation and QGOpt~\cite{10.21468/SciPostPhys.10.3.079} for updating tensors.

\subsection{Spin $S=1/2$ tetramer model}
\subsubsection{Hamiltonian and an exact ground state}
The Hamiltonian for the spin-$1/2$ tetramer model, which is a four-legs ladder model, is given by 
\begin{equation}\label{eq:tetramer_ham}
    \mathcal{H}=\sum_{x=1}^L\left(J \sum_{y=1}^4 \boldsymbol{s}_{x, y} \cdot \boldsymbol{s}_{x, y+1}+J^{\prime} \boldsymbol{S}_x \cdot \boldsymbol{S}_{x+1} \right)~
\end{equation}
with the total spin $\boldsymbol{S}_x=\sum_{y=1}^{4} \boldsymbol{s}_{x, y}$ of local tetramer under the torus boundary condition, $\boldsymbol{s}_{L+1, y}=\boldsymbol{s}_{1, y}$ and $\boldsymbol{s}_{x, 5}=\boldsymbol{s}_{x, 1}$, where $\boldsymbol{s}_{x, y}=(s^{\rm x}_{x,y},s^{\rm y}_{x,y},s^{\rm z}_{x,y})$ indicates the $S=1/2$ operator at coordinates $(x,y)$.
The coefficients $J$ and $J'$ represent the couplings in the intra-tetramer and the inter-tetramer interactions, respectively.

In this model, it is known that there is the tetramer singlets phase in range of $J'/J \lessapprox 0.628$ in the thermodynamic limit~\cite{doi:10.7566/JPSJ.87.123703}.
Until it deviates from this phase, the exact ground state forms a direct product of tetramer singlets. 

\subsubsection{Numerical settings}
To practically demonstrate, we applied our algorithm for Eq.~\eqref{eq:tetramer_ham} in case of $L=4$, $J'/J$ values of $0.0$, $0.741$, and $0.7411$, respectively.
The choice of $J'/J$ is based on the results obtained by the Lanczos method, which indicates that the exact ground state varies from the product state of tetramer singlets within the interval $0.741 < J'/J < 0.7411$ for Eq.~\eqref{eq:tetramer_ham} with $L=4$.
Although the ground states remain the same until breaking the tetramer singlets, as the $J'/J$ ratio increases and the frustration between interactions increases, this problem is expected to become more difficult as the energy spectrum becomes more dense at lower energies.

Here, we prepared $N_R=8$ replicas; each assigned a $\beta$ equally divided within $[\beta_{\min}, \beta_{\max}] = [6.0, 16.0]$, and we applied the following procedure for all replicas in parallel.
Firstly, we initialized the MERA state with random isometric tensors and updated them with $N_L = 2500$, $\delta = 10^{-14}$, and $\{ \eta_{\textrm{init}}, \eta_{\textrm{end}} \} = \{ 0.1, 0.0001 \}$.
After that, we executed our algorithm described in Sec.~\ref{sec:Method} and repeated this process ten times.
At each step 2 in Table.~\ref{Table_I}, we employed $N_L = 2500$ and $\delta = 10^{-12}$.
At each step 5, we used $N_L = 2500$ and $\delta = 10^{-14}$, and we performed the replica exchange in accordance with Eq.~\eqref{eq:replica exchange}.
We also used pruning parameters $\delta_{\textrm{structure}} = 10^{-1}$ and $\delta_{\textrm{replica}} = 10^{-2}$, respectively.
Note that in the steps 3 and 5, we adopted the learning rate
\begin{equation}
    \{ \eta_{\textrm{init}}, \eta_{\textrm{end}} \} = \{ (E - E_{\textrm{gs}}) \times 10^{-1}, (E - E_{\textrm{gs}}) \times 10^{-4} \}
\end{equation}
to prevent jumping the variational energy from the previous value due to a large learning rate.
In general, we cannot access the exact ground energy with a large system, so in this case, we could set $\eta$ with a constant number or approximate ground-state energy given by other numerical methods.

\subsubsection{Results}
Firstly, we will show an empirical result that the tetramer singlets state can be represented by a rebuilt structure from the 1D binary MERA, which consists of four physical bare sites with several disentanglers and isometries.
The Fig.~\ref{fig:test_tetramer} (a) represents the iteration of tensor-update dependence of relative errors 
\begin{equation}
\Delta = (E - E_{\text{gs}})/E_{\text{gs}}~
\end{equation}
between the variational energy $E$ and the exact ground energy $E_{\text{gs}}$ for Eq.~\eqref{eq:tetramer_ham} with $L = 4$, $J = 1.0$, and $J' = 0$.
Then, we employed two TNs as shown in Fig.~\ref{fig:test_tetramer} (b), where both TNs can go back and forth with each other by sequentially performing proposed recombination operations on the local structure.
In this experiment, all tensors were initialized randomly, meeting isometric conditions, and then we updated all of them with $N_L = 2500$, $\delta = 10^{-14}$, and $\{ \eta_{\textrm{init}}, \eta_{\textrm{end}} \} = \{ 0.1, 0.0001 \}$.
We confirm that the tetramer singlet structure can exactly represent the tetramer singlet state in numerical precision.

\begin{figure}[!t]
  \centering
  \includegraphics[clip, width=3.4in]{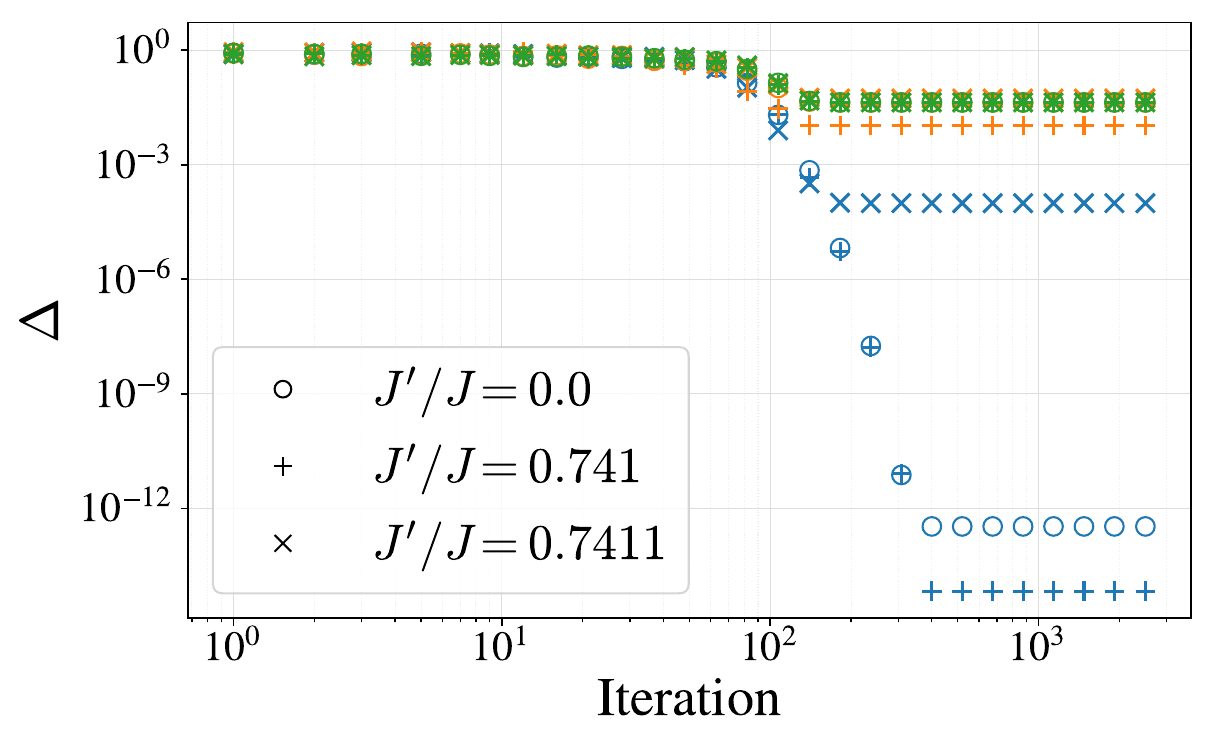}
  \caption{The iteration dependence of relative errors $\Delta$ for the binary MERA (green symbols), optimized structure without the heat-bath (orange symbols) and with the heat-bath and the replica exchange (blue symbols) for several $J'/J$.
  }
  \label{fig:tetramer_energy}
\end{figure}

Figure \ref{fig:tetramer_energy} shows iteration dependencies of the relative errors $\Delta$ for $J'/J = 0.0$, $0.741$, and $0.7411$ respectively, for the MERA state and reconstructed two structures after applying our algorithm; the instance without the heat-bath $(\beta \rightarrow \infty)$ and those showing minimum energy error within all replicas.
In this figure, to show the differences in results between obtained structures, we initialized tensors randomly, meeting isometric conditions, and updated all of them with $N_L = 2500$, $\delta = 10^{-14}$, and $\{ \eta_{\textrm{init}}, \eta_{\textrm{end}} \} = \{ 0.1, 0.0001 \}$.
In cases where the ratio $J'/J$ are $0.0$ and $0.741$, it is evident that our algorithm incorporating replica exchange converges nearly to the exact ground energy, on the other hand at the instance without the heat-bath only minimal improvements are observed.
It supports empirical evidence that replica exchange and heat-bath methods with finite temperatures could increase the potential to find instances that close the targets.
Note that the number of times firstly converging to the exact ground energy was observed, the fourth time in $J'/J=0.0$ and the ninth time in $J'/J=0.741$, out of the ten repetitions of our algorithm.
This could be because the variational energy landscape becomes more complex as the gap between the ground and first excited states is smaller for the increase of $J'/J$.

\begin{figure}
  \centering
    \includegraphics[width=2.5in]{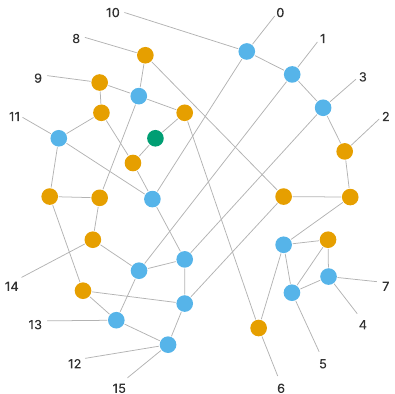}
    \caption{The diagram of the structure showing minimum energy error value in case of $J'/J = 0.0$.
    The numbers assigned to bare spins represent indices of Hamiltonian.}
    \label{fig:tetramer_stuctures}
\end{figure}

To investigate the obtained structures in detail, we illustrate the structure showing minimum energy error $\Delta$ in $J'/J = 0$ (Fig.~\ref{fig:tetramer_energy}) as Fig.~\ref{fig:tetramer_stuctures}.
It suggests that the obtained TN is not a trivial structure composing tetramer singlets, as shown in the right panel of Fig.~\ref{fig:test_tetramer} (b).
We also found that some exact tetramer singlets are formed through the optimization process, even if inter-tetramers are not completely disentangled in the sense of structure.
This behavior is one of the interesting phenomena that indirectly suggests the TN possesses more than enough degrees of freedom to form a tetrameric singlet state.

In the case of $J'/J = 0.7411$, the exact ground state is associated with $S=2$ Haldane phase.
We did not obtain exact ground energy as in the singlet phase case, but we reduced the error from $4.30 \times 10^{-2}$ at the initial state to $9.88 \times 10^{-5}$ as shown in Fig.~\ref{fig:tetramer_energy}. 

\subsection{$S=1/2$ one-dimensional Random XY chain}
\subsubsection{Hamiltonian}

The $S=1/2$ 1D random XY chain with periodic boundary condition is given by
\begin{equation}\label{eq:XY_hamiltonian}
    \mathcal{H}= \sum_{i=1}^N J_i \left[s_i^{\rm x} s_{i+1}^{\rm x}+s_i^{\rm y} s_{i+1}^{\rm y}\right]~,
\end{equation}
where $N$ is the system size, $(s_i^{\rm x},s_i^{\rm y})$ denotes $S=1/2$ operators for $i$th site, $s_{i+N}^{\rm x(y)} = s_{i}^{\rm x(y)}$, and the random coupling parameters $\{ J_i \in \mathbb{R} \}$ are allowed to vary in the range $[0,1]$ with uniform distribution.
It is known that spacial averaged entanglement entropy of the exact ground state for the Hamiltonian obeys the logarithmic scaling~\cite{PhysRevLett.93.260602}.

\subsubsection{Numerical settings}
We demonstrated our algorithm for two initial structures: binary MERA and sub-optimal structures obtained using the SDRG method for ER proposed in Ref.\cite{PhysRevB.96.155136}, for system sizes $N = 8$ and $16$.
The latter is explained in more detail in Appendix.~\ref{appendix: ER_SDRG}, is referred to in this paper as ER-SDRG.
For $N=8$, the internal degrees of freedom on tensors of MERA and ER-SDRG are $175$ and $172$, and for $N=16$ they are $399$ and $392$, respectively.
Thus, the fact that the difference in internal degrees of freedom between the two structures consistently remains within 3\% facilitates discussions on how the structure of the initial TN influences the performance of our algorithm.
For both structures, we prepared one replica with $\beta \rightarrow \infty$ for 50 disorder realizations with randomly varying coupling strengths.
In all prepared structures, we initialized tensors randomly maintaining isometric conditions and updated them with $N_L = 5000$, $\delta = 10^{-12}$ and $\{ \eta_{\textrm{init}}, \eta_{\textrm{end}} \} = \{ 0.1, 0.0001 \}$.
After that, we repeated applying the proposed procedure five times.
The total number of iterations was set to $N_L = 5000$ for both the steps 2 and 5, and the thresholds were set to $\delta = 10^{-6}$ and $10^{-12}$ at the steps 2 and 5, respectively.
About learning rate $\eta$ in the steps 3 and 5 were defined the same as in the tetramer model.

\subsubsection{Results}
\begin{figure}[tb]
  \centering
  \includegraphics[clip, width=3.2in]{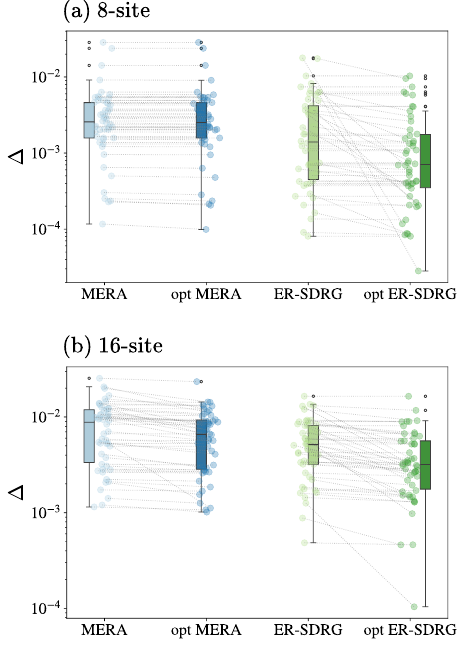}
  \caption{The benchmark for the relative energy errors in MERA and ER-SDRG structures. "opt $\circ$" represents the results after applying our method, starting with $\circ$ as the initial structure. The dashed lines connecting the circles before and after the optimization indicate the change in relative errors for each disorder configuration.}
  \label{fig:XY_energy}
\end{figure}
\begin{figure}[tb]
  \centering
  \includegraphics[clip, width=3.2in]{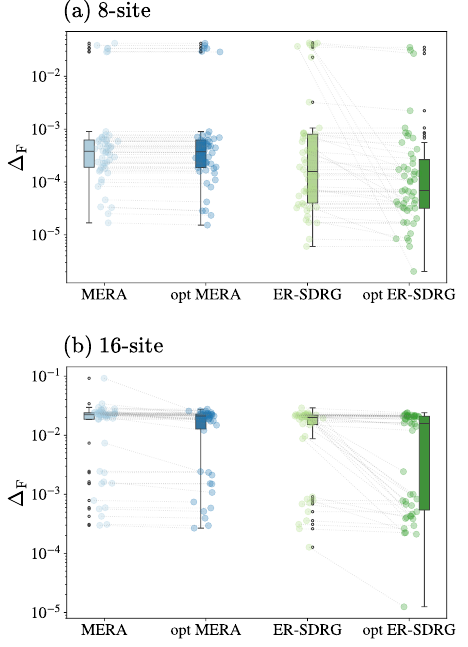}
  \caption{The benchmark for the infidelity at each bare bond in MERA and ER-SDRG structures. The explanation of this type of figure is already described in Fig.~\ref{fig:XY_energy}.}
  \label{fig:XY_infidelity}
\end{figure}
The panel in Fig.~\ref{fig:XY_energy} compares the relative errors $\Delta$ from the exact ground state energy $E_{\text{gs}}$ in the case of system sizes $N=8$ and $16$ across four different structures.
The horizontal axis represents structural differences, where "opt $\circ$" denotes an optimized $\circ$ network using our method.
The result indicates that our method can lead to better solutions, with a lower average $\Delta$ from the initial structure in both the MERA and ER-SDRG.
Specifically, we observed the decreasing rate of the relative errors
\begin{equation}\label{eq:decrease ratio}
    R_{\textrm{TN}} = \left( 1 - \frac{\Delta({\textrm{opt TN}})}{\Delta({\textrm{TN}})} \right) \times 100~[\%]
\end{equation}
with ${\rm TN} \in \{ \textrm{MERA,~ER-SDRG} \}$ in both 8-site and 16-site configurations as follows:
\begin{center}
    \begin{tabular}{|c|c|c|}
       \multicolumn{3}{c}{8-site} \\ \hline
       TN & MERA & ER-TN \\ \hline\hline
       mean & 1.37\% & 24.9\% \\ \hline
       max & 15.1\% & 99.6\% \\ \hline
    \end{tabular}
    ~ and ~ 
    \begin{tabular}{|c|c|c|}
       \multicolumn{3}{c}{16-site} \\ \hline
       TN & MERA & ER-TN \\ \hline\hline
       mean & 15.2\% & 26.0\% \\ \hline
       max & 79.0\% & 91.7\% \\ \hline
    \end{tabular}~.
\end{center}
From this analysis, our algorithm succeeded more for ER-SDRG, and this suggests that our method works more powerfully when utilizing the TN design
methods for non-uniform systems as a preprocessing step.

Next, we compare the infidelity per site 
\begin{equation}
\Delta_{\textrm{F}} = 1 - |\braket{\Psi}{\Psi_{\text{gs}}}|^{1/N}~,
\end{equation}
between the variational state $\ket{\Psi}$ and the exact ground state $\ket{\Psi_{\text{gs}}}$ in Fig.~\ref{fig:XY_infidelity}.
In random systems, fidelity could be significantly boosted with a slight improvement in energy because many low-energy states tend to emerge more than in uniform systems.
We observed the decrease ratio in Eq.~\eqref{eq:decrease ratio} as follows:
\begin{center}
    \begin{tabular}{|c|c|c|}
       \multicolumn{3}{c}{8-site} \\ \hline
       TN & MERA & ER-TN \\ \hline\hline
       mean & 1.59\% & 25.2\% \\ \hline
       max & 15.4\% & 99.994\% \\ \hline
    \end{tabular}
    ~and~
    \begin{tabular}{|c|c|c|}
       \multicolumn{3}{c}{16-site} \\ \hline
       TN & MERA & ER-TN \\ \hline\hline
       mean & 10.4\% & 30.9\% \\ \hline
       max & 76.2\% & 98.3\% \\ \hline
    \end{tabular}~.
\end{center}
Excluding the results from the 16-site MERA, the decreasing rates in fidelity error per site were greater than those in relative energy error.

\begin{figure}[H]
  \centering
  \includegraphics[clip, width=3.4in]{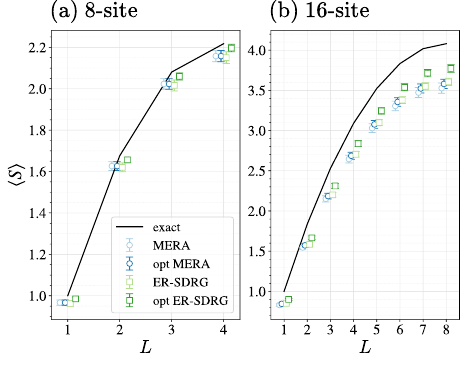}
  \caption{The results are averaged over all possible subsystem sizes $L$ for 50 disorder configurations. We plot the results for different structures at each $L$, which are slightly shifted along the $x$-axis to improve visibility.}
  \label{fig:XY_EE}
\end{figure}

Finally, in Fig.~\ref{fig:XY_EE}, we show the average entanglement entropy $\expval{S}$ of any possible subsystem size $L$ for 50 disorder configurations, where $S = -{\rm Tr}[\rho_{\rm A} \log_2(\rho_{\rm A})]$ with $\rho_{\rm A} = \Tr_{\rm B}{[ \ketbra{\Psi} ]}$ is the entanglement entropy defined when the $N$-site system is divided into a subsystem A that includes $L$ sites of the system and a subsystem B that contains the remaining sites. 
Here, $\Tr_{\rm B}$ means the partial trace for the physical degrees of freedom associated with subsystem B.
Our method, which refers to variational energy, also improved entanglement entropy.
Similar to $\Delta$ and $\Delta_{\textrm{F}}$, our method against ER-SDRG apparently yielded superior results compared to the MERA state.
In this experiment, we did not impose U(1) symmetry on each tensor, and it is interesting to see if our method will produce further improvements when considering symmetry.
This is because it is generally known that tensor network methods with low bond dimensions tend to break the symmetry of the system and favor states with low entanglement.

\section{Conclusion}\label{sec:Conclusion}
In this paper, we have proposed the method for structural search of ER-TN to approach the ground state of quantum many-body systems.
While existing references utilize SDRG to construct ER-TN for disordered systems, our study explores the reconstruction of local structures.
Compared to TTNs, which can directly evaluate entanglement entropy between two subsystems based on singular values at the canonical center~\cite{PhysRevResearch.5.013031}, ER-TN cannot perform the evaluation straightforwardly.
This is because ER-TN do not always allow defining canonical centers because of internal loops.
Thus, we focused on rearranging two local tensors in ER-TNs by referring to the optimized variational energy, which improves the overall structures.
Our study demonstrated the potential of our approach in optimizing structures for both the spin $S=1/2$ tetramer and random XY models.

In the spin $S=1/2$ tetramer model, we encountered a problem where our algorithm does not converge to the exact target state when continually adopting the structure with minimum variational energy.
To address this problem, we introduced the concept of the heat-bath method and the replica exchange.
When applied iteratively with replica exchange, our algorithm proved effective in scenarios with strong inner-tetramer interactions until the direct product of tetramer singlets state is disrupted.
Furthermore, it successfully reduce the relative error by $R = 99.8$ after breaking the tetramer singlets state.

For the 1D random XY model, we applied our method to both MERA and TNs obtained by SDRG to verify the high accuracy of the variational state.
Our results showed energy, fidelity, and entanglement entropy improvements for both initial structures.
Our method's efficacy is more pronounced for the latter structure, highlighting the importance of providing suitable initial structures.

Our research concept shares similarities with previous studies updating the connectivity of TN structures.
Automatic quantum circuit encoding (AQCE)~\cite{shirakawa2021automatic} involves the process of reconnecting each quantum gate sequentially in quantum circuits, which can be classified as ER-TN.
AQCE is an algorithm that transforms any quantum state into a quantum circuit by iteratively inserting optimal gates at the best positions in each circuit depth.
AQCE always focuses on one two-qubit gate to decide connectivity; conversely, our method differs by targeting adjacent gate pairs and introducing the choice for stochastic selection of local structures.
Our method, of course, can be used with such algorithms to address problems for quantum computing.

Expanding the bond dimension and applying it to larger systems are essential steps to enhance the applicability of our method.
Contraction techniques for TN with loop structures will be crucial for achieving this goal~\cite{10.21468/SciPostPhys.15.6.222,PhysRevB.99.195105}.
Moreover, given our algorithm's focus on local regions of the network, it will be effective to explore the parallelization of each step as seen in the real-space parallel algorithm for TN~\cite{PhysRevB.87.155137,doi:10.7566/JPSJ.87.074005,sun2023improved}. 
Another area of focus will be refining tensor update methods to improve both efficiency and accuracy.
We aim to develop algorithms capable of algebraically decomposing high-rank tensors into two tensors to follow each structure, similar to the two-tensor DMRG approach.

\begin{acknowledgments}
This work is partially supported by,
KAKENHI Grant Numbers JP22H01171, JP21H04446,
and a Grant-in-Aid for Transformative Research Areas titled "The Natural Laws of Extreme
Universe---A New Paradigm for Spacetime and Matter from Quantum Information"
(KAKENHI Grant Nos. JP21H05182, JP21H05191) from JSPS of Japan.
We acknowledge support from MEXT Q-LEAP Grant No.
JPMXS0120319794, and JST COI-NEXT No. JPMJPF2014.
R.W. would like to thank T. Hikihara and K. Fujii for useful discussions and was supported by the $\Sigma$ Doctoral Futures Research Grant Program from Osaka University.
H.U. was supported by the COE research grant in computational science from Hyogo
Prefecture and Kobe City through Foundation for Computational Science.
We are also grateful for the allocation of computational resources of the HOKUSAI BigWaterfall supercomputing system at RIKEN and SQUID at the Cybermedia Center at Osaka University.
\end{acknowledgments}

\appendix
\section{Specific algorithm for selection local two tensors}\label{appendix: our_algorithm}

\begin{algorithm}[H]
\caption{Select local tensors}
\label{alg:choice edge}
\begin{algorithmic}[1]
\Require integer $j \in [0, N_{\rm E}]$, set of pair of integers $\mathcal{E}=\{ e_i \in [1, N_{\rm T}]^{\otimes 2} \}$, set of binary $\mathcal{F}=\{ f_i \in \{0,1\} \}$, set of integer $\mathcal{D}=\{d_i \in [1,N_{\rm T}]\}$ for $1 \leq i \leq N_{\rm E}$
\Ensure integer $j_{\textrm{new}} \in [1, N_{\rm E}]$
    \If{$j$ is 0} \label{alg:void}
        \Statex \Comment $j=0$ means our algorithm is at the first selection.
        \State $Q \leftarrow \{i \in [1, N_{\rm E}] \mid f_i = 0\}$
        \State $d_{\textrm{max}} \leftarrow \max_{q \in Q}{\{d_q\}}$
        \State $R \leftarrow \{q \in Q \mid d_q = d_{\textrm{max}} \}$
    \Else
        \Statex \Comment $j(\in [1,N_{\rm E}])$th edge was selected in the previous step 1.
        \State $f_j \leftarrow 1$ \label{alg:f(o)}
        \State $P \leftarrow $ \Call{adjacent\_edge\_indices}{$j$, $\mathcal{E}$}
        \Statex \Comment The function \Call{adjacent\_edge\_indices}{$i$, $\mathcal{E}$} returns indices of edge in $\mathcal{E}$, where they connect with two tensors which are indicated $e_i$ except for $i$ itself.
        \State $Q \leftarrow \{p \in P \mid f_p = 0 \}$
        \If{$Q$ is $\varnothing$}
        \label{alg:exception}
            \State Execute the same procedure as in line 2.
        \EndIf
        \State Execute the same procedure as in lines 3-4.
    \EndIf
    \State {$j_{\textrm{new}}$ is sampled from $R$ \label{alg:last}}
\State \Return $j_{\textrm{new}}$
\end{algorithmic}
\end{algorithm}

In Sec.~\ref{sec:Method}, we briefly described how to select local tensors in our algorithm.
To ensure clarity and thoroughness, we will further explain the methods we adopted using the specific procedure to select local tensors shown in Algorithm \ref{alg:choice edge}, which is used in step 1 of Table.~\ref{Table_I}.
This procedure can reconstruct all possible local structures only once within step 4 in Table.~\ref{Table_I}, while prioritizing the strategies described in Sec.~\ref{sec:Method}.
It should be noted that there may be a more optimal approach that would improve the performance of our overall network recombination algorithm.

\section{Entanglement Renormalization for disordered systems}\label{appendix: ER_SDRG}
Establishing a sub-optimal structure in advance is crucial for our method.
We employed the method proposed in Ref.~\cite{PhysRevB.96.155136} to prepare an initial structure for the disordered systems.
Here, we provide an overview of this method.

\begin{figure}[H]
  \centering
  \includegraphics[clip, width=2.8in]{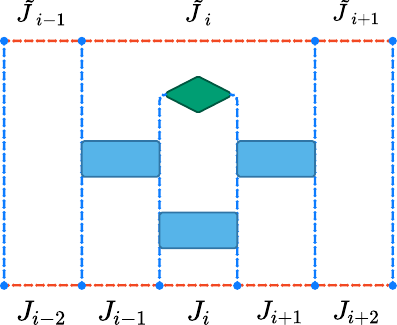}
  \caption{A Schematic diagrams illustrating the construction of the optimal structure by the method mentioned in Ref.~\cite{PhysRevB.96.155136}}
  \label{fig:ER-SDRG}
\end{figure}

Unlike the local reconstruction approach, this algorithm constructs the structure by referencing the coupling constants of the Hamiltonian. 
For the Hamiltonian, exemplified in Eq.~\eqref{eq:XY_hamiltonian}, this method approximates a state for the pair of spins with the strongest coupling $J_i$ as the singlet.
Then, using the second-order perturbation theory, we determine the effective coupling
\begin{equation}
    \tilde{J}_i = \frac{J_{i-1}J_{i+1}}{J_i}~,
\end{equation}
between the spins on either side, as depicted in Fig.~\ref{fig:ER-SDRG}.
This network contains two kinds of local structures: $\{u,u\}$ (Fig.~\ref{fig:local_structures}(d)) and $\{u,t\}$ shown in Fig.~\ref{fig:local_structures}(e).
Note that, while Ref.~\cite{PhysRevB.96.155136} fixing $t$ tensors as the singlet state and impose $u$ tensors are updated while meeting $U(1)$ symmetry, our method updates all tensors without this symmetry.

\bibliography{main}
\end{document}